\documentclass[prd,aps,amsmath,amssymb,nofootinbib]{revtex4}

\usepackage{graphicx}
\usepackage{subfigure}
\usepackage{bm}
\usepackage{amsmath}
\usepackage{array}
\usepackage{multirow}
\usepackage{float}
\usepackage{color}
\graphicspath{{figs/}}

\begin{document}

\title{Investigating the possibility of a turning point in the dark energy equation of state}

 \author{Yazhou Hu$^{1,2}$, Miao$^{3,1}$, Xiao-Dong Li$^{4}$ and Zhenhui Zhang$^{1,2}$}
%
  \affiliation{1) State Key Laboratory of Theoretical Physics, Institute of Theoretical Physics, Chinese Academy of Sciences, Beijing, 100190}
  \affiliation{2) Kavli Institute for Theoretical Physics China, Chinese Academy of Sciences, Beijing, 100190}
  \affiliation{3) School of Astronomy and Space Science, Sun Yat-Sen University, Guangzhou, 510275, China}
  \affiliation{4) Korea Institute for Advanced Study, Hoegiro 87, Dongdaemun-Gu, Seoul, 130-722, Korea}


\begin{abstract}
We investigate a second order $parabolic$ parametrization, 
$w(a)=w_t+w_a(a_t-a)^2$,
which is a direct characterization of a possible $turning$ in $w$.
The cosmological consequence of this parametrization is explored by using the observational data of the SNLS3 type Ia supernovae sample,
the CMB measurements from WMAP9 and Planck,
the Hubble parameter measurement from HST,
and the baryon acoustic oscillation (BAO) measurements from 6dFGS, BOSS DR11 and improved WiggleZ.
We found the existence of a turning point in $w$ at $a\sim0.7$ is favored at 1$\sigma$ CL.
In the epoch $0.55< a< 0.9$, $w<-1$ is favored at 1$\sigma$ CL,
and this significance increases near $a=0.8$, reaching a 2$\sigma$ CL.
The parabolic parametrization achieve equivalent performance to the $\Lambda$CDM and Chevallier-Polarski-Linder (CPL) models
when the Akaike information criterion was used to assess them.
Our analysis shows the value of considering high order parametrizations when studying the cosmological constraints on $w$.
\end{abstract}

\pacs{95.36.+x, 98.80.–k}
\maketitle

\section{Introduction}

Cosmic acceleration was discovered in 1998 \citep{Riess1998,Perlmutter1999}
and has now been confirmed by several independent observations.
This discovery implies the possible existence of a new energy component ``dark energy''.
Although widely investigated in the last decade \citep{SS2000,PR2003,FTH2008,Li2011},
the nature of dark energy still remains a mystery.

Observational explorations are of essential importance to understanding dark energy.
Currently, most powerful probes of the cosmic expansion history
include the type Ia supernovae (SNIa) and baryon acoustic oscillation (BAO) measurements.
SNIa as standard candles provide direct measurements of the luminosity distance as a function of redshift,
while the BAO features in the clustering of galaxies as standard rulers
provide robust measurements of the angular diameter distance-redshift relation
when calibrated by the high redshift cosmic microwave background (CMB) observations
\citep{PeeblesBAO,Eisenstein:1997ik}.

Recently, there are two sets of BAO measurements released with unprecedented precision.
Recently, the Baryon Oscillation Spectroscopic Survey (BOSS)
reported a 2\% constraint on the volume averaged cosmic distance
$D_{\rm V}(z=0.32)\left({r^{\rm fid}_{\rm d}}/{r_{\rm d}}\right)=1264\pm25 {\rm Mpc}$
measured from the Data Release 11 (DR11) LOWZ sample,
and a $1\%$ result $D_{\rm V}(z=0.57)\left({r^{\rm fid}_{\rm d}}/{r_{\rm d}}\right)=2056\pm20 {\rm Mpc}$ measured from the DR11 CMASS sample \citep{dr11}.
The latter measurement is the most precise distance constraint ever achieved by galaxy surveys.
Later, Kazin et al. \cite{impwigz} presented improved distance measurements from the WiggleZ Dark Energy Survey at $z=0.44$, 0.6 and 0.73.
By employing a reconstruction technique to correct the baryonic oscillations for the smearing caused by galaxy peculiar velocities \citep{recon},
they achieved significant improvements in the measurements,
equivalent to those expected from surveys with up to 2.5 times the volume of WiggleZ.
Therefore, it is noteworthy to investigate the constraints on dark energy from these precise BAO data.

It is widely believed that dark energy can be characterized by its equation of state (EOS) $w\equiv p/\rho$,
customarily treated as a general function of the scale factor $a$.
To determine $w(a)$ from data, one could solve the dynamical equation for a particular theory,
but the result of this approach depends on the adopted theory.
Instead, for generality, various parametrizations of $w(a)$ are used.
Among the numerous parametrizations proposed thus far (see \cite{Li2011} and the references therein),
a most popular one is the Chevallier-Polarski-Linder (CPL) parametrization \citep{CPLA,CPLB}
\begin{eqnarray}\label{eq4}
w(a)=w_0+w_a(1-a) =w_0+w_a\frac{z}{1+z}.
\end{eqnarray}
This model characterizes dark energy by two free parameters.
Parameter $w_0$ determines the present-day value of $w$,
while parameter $w_a$ characterizes the first-order derivative of $w$ with respect to $a$.
The CPL parametrization has many obvious advantages, for instance,
a manageable parameter space, the bounded behavior at high redshift,
good sensitivity to observational data,
and the ability to accurately reconstruct many dark energy theories (see \cite{CPLB} for details).
Thus, it was widely used in the literature to reconstruct dark energy from observational data.

Despite the various merits, the ability of CPL parametrization in reconstructing dark energy
is still limited because of its simple, linear form.
It cannot characterize a $w$ with high-order features, 
such as one or more extreme points, 
oscillations or fast transitions at some epoch,
while such behaviors appear in many dark energy theories -
such as quintessence \citep{BCN2000}, modified gravity \citep{HS2007,Star2007,AB2007},
coupled dark energy \citep{Baldi2012,HDEB} models.
To study these models, high-order parametrizations are needed.
In addition, numerical studies of dark energy based on observations also have the demand for considering high-order dark energy parametrizations.
For example, Alam et al. \cite{Alam2003} adopted a high-order parametrization of dark energy density and found that a time-dependent dark energy is favored by the supernovae data.
Zhao et al. \cite{Gongbo Zhao} applied a non-parametrization technique to reconstruct the dark energy EOS,
and showed that a dynamical dark energy is mildly favored over $\Lambda$CDM.
Finally, high-order dark energy parametrizations are required studying since there is no reason for us to limit the anlaysis at the linear order.

Various high-order dark energy parametrizations have been considered in many works (see \cite{Li2011} and the references therein).
In this work, for the purpose of studying the possibility of a $turning$ point (or equivalently, an extreme point) in $w$,
we propose a $parabolic$ form
\begin{eqnarray}\label{eq6}
w(a)=w_t+w_a(a_t-a)^2 =w_t+w_a\left(a_t-\frac{1}{1+z}\right)^2.
\end{eqnarray}
Seemingly, this parametrization has three free parameters,
enabling a turning of $w(a)$ at $a=a_t$.
The extreme value $w(a=a_t)$ and the sharpness of the tuning are characterized by $w_t$ and $w_a$, respectively.
Like CPL, the parabolic model also has bounded behavior at high redshift.
Mathematically, this model is equivalent to a second order expansion $w(a) = w_0 + w_1 a + w_2a^2$,
but it has apparent advantages that the parameters $w_t$, $a_t$, $w_a$ are direct characterizations of a turning point in $w(a)$.
In addition, since $a_t$ is not necessarily confined to the region [0,1],
the model is also able to describe those $w(a)$s which do not have turning points in the past epoch.

In this work we do not adopt the non-parametric method \citep{H2010,SKL2012,SCS2012,C2012,Gongbo Zhao,WD2014}.
The reason is that the parabolic parametrization is a more direct characterization of a turning in $w$.
Also, we would not consider expansions of $w(a)$ at third or higher orders, since a second order expansion should be sufficiently comprehensive.
Linder et al. \cite{LH2005} argued that two dark energy parameters are enough and even future experiments can not put accurate constraints on more parameters.
A more exhaustive analysis done by Sarkar et al. \cite{Sarkar2008} showed that next-generation dark energy surveys
may be able to constrain three or more independent parameters of $w$ to better than 10\%.
Regardless, Eq. (\ref{eq6}) shall be sufficiently comprehensive given the current observational data.

\section{Methodology}

We explore the cosmological constraints on dark energy via the parabolic model.
For comparison, we will also present the fitting results of the $\Lambda$CDM and CPL models.
For simplicity, in our analysis we assume a flat universe with $\Omega_k=0$.
So the expansion rate of the universe $H(z)$ (i.e. the Hubble parameter) is given by
\begin{equation}\label{eq1}
H(z)=H_0\left[\Omega_{m}(1+z)^3+\Omega_{r}(1+z)^4+(1-\Omega_m-\Omega_r)f(z)\right]^{1/2},
\end{equation}
where $H_0$ is the current value of the Hubble parameter,
$\Omega_m$($\Omega_r$) is the current ratio between the matter(radiation) density and the critical density of the Universe.
Contribution from the radiation component is characterized by $\Omega_r=\Omega_\gamma(1 + 0.2271N_{\rm eff})$,
where $\Omega_\gamma=2.469\times 10^{-5}h^{-2}$ ($\gamma$ denotes photon) as a result of $T_{\rm cmb}=2.725$ K,
and the effective number of neutrino species $N_{\rm eff}$ assumes its standard value 3.04 \citep{WMAP7}.
The dark energy density function takes the form $f(z)=\exp[3\int_0^z dz'(1+w(z'))/(1+z')]$.

We use the most recent observational data to perform $\chi^2$ analyses and explore the parameter space of the model.
Data used in our analysis include:
\begin{itemize}
\item
The SNLS3 (Supernova Legacy Survey 3-year) combined sample~\citep{SNLS3A,SNLS3B}, consisting of 472 SNIa,
combining the results of two light-curve fitting codes SiFTO \citep{SiFTO} and SALT2 \citep{SALT2}.
We follow the procedure of \cite{Zhenhui Zhang} and perform a complete analysis of the systematic errors.
The SNLS3 $\chi^2$ function takes the form
\begin{equation}
\chi^2_{SNLS3}=\Delta \overrightarrow{\bf m}^T \cdot {\bf C}^{-1} \cdot \Delta \overrightarrow{\bf m},
\end{equation}
where {\bf C} is a $472 \times 472$ covariance matrix capturing the statistic and systematic uncertainties,
and $\Delta {\overrightarrow {\bf m}} = {\overrightarrow {\bf m}}_B - {\overrightarrow {\bf m}}_{\rm mod}$ is a vector of
model residuals of the SNIa sample,
with $m_B$ the rest-frame peak $B$ band magnitude of the SNIa and
$m_{\rm mod}$ the predicted magnitude of the SNIa, given by
\begin{equation}\label{SNchisq}
m_{\rm mod} = 5\log_{10} \mathcal{D}_L-\alpha(s-1)+\beta \mathcal{C} + \mathcal{M},
\end{equation}
where $\mathcal{D}_L$ is the Hubble-constant free luminosity distance,
the stretch $s$ is a measure of the shape of SN light-curve, $\mathcal{C}$ is the color measure for the SN,
and $\alpha$, $\beta$ are two nuisance parameters characterizing the stretch-luminosity and color-luminosity relationships, respectively.
Following \cite{SNLS3A}, we treat $\alpha$ and $\beta$ as free parameters of $\chi^2$ function.
Note that the covariance matrix $\mathcal{C}$ depends on $\alpha$ and $\beta$,
so it is reconstructed and inverted every time when the values of $\alpha$ and $\beta$ are varied.
The nuisance quantity $\mathcal{M}$ is a combination of the absolute magnitude of a fiducial SNIa and the Hubble constant.
We marginalize it following the complex formula in the Appendix C of \cite{SNLS3A}.
The host-galaxy information is included by splitting the samples into two parts and allowing the absolute magnitude to be different between these two parts.
More information on SNLS3 can be obtained elsewhere \cite{SNLS3A}.

\item
We use the CMB measurements from the WMAP 9-year \citep{WMAP9} and Planck first year \citep{PlanckA,PlanckB} observations.
The WMAP9 and Planck ``distance priors'' are provided in the analysis~\cite{Yun Wang},
including the baryon component $\omega_b\equiv \Omega_b h^2$, the ``acoustic scale''
$l_a\equiv{\pi r(z_*)/ r_{\rm s}(z_*)}$, and the ``shift parameter'' $R\equiv{\sqrt{\Omega_m H_0^2} \,r(z_*)}$,
where $z_*$ is the redshift to the photon-decoupling surface \citep{Hu:1995en},
$r(z_*)$ is our comoving distance to $z_*$,
and $r_{\rm s}(z_*)$ is the comoving sound horizon at $z_*$.
The distance priors provide an efficient summary of the CMB data inregards to dark energy constraints \citep{Li2011}.

\begin{table*} \caption{BAO measurements used in this analysis. }
\begin{center}
\label{Table0}
\begin{tabular}{ccc}
  \hline\hline
  Survey      ~&~$z_{\rm eff}$   & Constraint    \\
  \hline
  6dFGS       ~&~ 0.106 & $r_{\rm s}/D_{\rm V}=0.336\pm0.015$            \\
  \hline
  BOSS, DR11  ~&~ 0.32 & $D_{\rm V}\left({r^{\rm fid}_{\rm d}}/{r_{\rm d}}\right)=1264\pm25 {\rm Mpc}$ $\ ^a$            \\
              ~&~ 0.57 & $D_{\rm A}\left({r^{\rm fid}_{\rm d}}/{r_{\rm d}}\right)=1421\pm20 {\rm Mpc}$ $\ ^a$           \\
              ~&~ 0.57 & $H\left({r_{\rm d}}/{r^{\rm fid}_{\rm d}}\right)=96.8\pm3.4 {\rm km s^{-1} Mpc^{-1}}$ $\ ^a$           \\
  \hline
  WiggleZ  ~&~ 0.44 & $D_{\rm V}\left({r^{\rm fid}_{s}}/{r_{\rm s}}\right)=1716\pm83 {\rm Mpc}$ $\ ^b$         \\
           ~&~ 0.60 & $D_{\rm V}\left({r^{\rm fid}_{s}}/{r_{\rm s}}\right)=2221\pm101 {\rm Mpc}$ $\ ^b$        \\
           ~&~ 0.73 & $D_{\rm V}\left({r^{\rm fid}_{s}}/{r_{\rm s}}\right)=2516\pm86 {\rm Mpc}$ $\ ^b$           \\
  \hline\hline
\end{tabular}
\end{center}
\leftline{\noindent$\ ^a$ $r^{\rm fid}_{\rm d}=153.19 {\rm Mpc}$ for BOSS DR11 distance priors.}
\leftline{\noindent$\ ^b$ $r^{\rm fid}_{s}=152.3 {\rm Mpc}$ for WiggleZ distance priors.}
\end{table*}

\begin{table*} \caption{Fitting results of the parabolic model.}
\begin{center}
\label{Table1}
\begin{tabular}{ccc}
  \hline\hline
  Dataset   &           SNLS3+Planck+BAO+HST & SNLS3+WMAP9+BAO+HST \\
  \hline
  $\Omega_{m}$ &  $0.285\pm 0.010 ^a$ &  $0.286\pm0.011 $    \\
  $w_t$    &   $-1.108^{+0.15}_{-0.16} $~~&    $-1.13^{+0.23}_{-0.21} $~                            \\
  $w_a$  &   $-0.3^{+1.0}_{-0.9} $ ~~& $-0.4^{+1.3}_{-1.0} $ \\
  $a_t$  & $0.75\pm0.65  $ ~~& $0.76^{+0.74}_{-0.75} $ \\
  \hline
  $\chi^2_{min}$  &   424.83 &~~       424.86   \\
  \hline\hline
\end{tabular}
\end{center}
\leftline{\noindent$\ ^a$ We list the mean value and 68\% limit.}
\end{table*}

\item
The BAO data used in our analysis are listed in Table \ref{Table0}.
They include the measurement of $r_{\rm s}/D_{\rm V}$ at $z=0.106$ from 6dFGS (6-degree Field Galaxy Survey)~\citep{Beutler},
the isotropic measurement of $D_{\rm V}/r_{\rm d}$ at $z=0.32$ from the BOSS DR11 LOWZ sample \citep{dr11},
the anisotropic measurement of $D_{\rm A}/r_{\rm d}$ and $Hr_{\rm d}$ at $z=0.57$ from the BOSS DR11 CMASS sample \citep{dr11},
and the improved measurements of $D_{\rm V}/{r_{\rm s}}$ 
at $z=0.44$, 0.60, 0.73 from the WiggleZ Dark Energy Survey \citep{impwigz}.
Here $r_{\rm d}$ is the comoving sound horizon at the ``drag'' epoch when the baryons are ``released'' from the drag of the photons \citep{Eisenstein:1997ik},
and $D_{\rm V}$ is a volume averaged distance indicator similar to the angular diameter distance $D_{\rm A}$ ~\citep{eisenstein.etc}.
Following similar previous studies \citep{Addison2013}, 
we consider the covariance between different BAO surveys to have a negligible effect on our statistical significance 
and thus ignore this effect in our analysis.
We would not use the distance measurement from the SDSS DR7 sample \citep{Padamanabhan} to avoid overlap with the BOSS DR11 LOWZ sample.

\item
We also use the direct measurement of the Hubble constant $H_0=73.8\pm 2.4 {\rm km/s/Mpc}$
from the supernova magnitude-redshift relation calibrated by the HST (Hubble Space Telescope)
observations of Cepheid variables in the host galaxies of eight type Ia supernovae~\citep{HSTWFC3}.
Here the 1$\sigma$ uncertainty includes known sources of systematic errors.
\end{itemize}
In the following context, we will use ``SNLS3'', ``CMB'', ``BAO'' and ``HST'' to
represent these four datasets. For the ``CMB'' data set, we will use ``Planck''
and ``WMAP9'' to represent the Planck and WMAP9 distance priors, respectively.

\begin{figure*}[htbp]
\centering
\begin{center}
\includegraphics[scale=0.8]{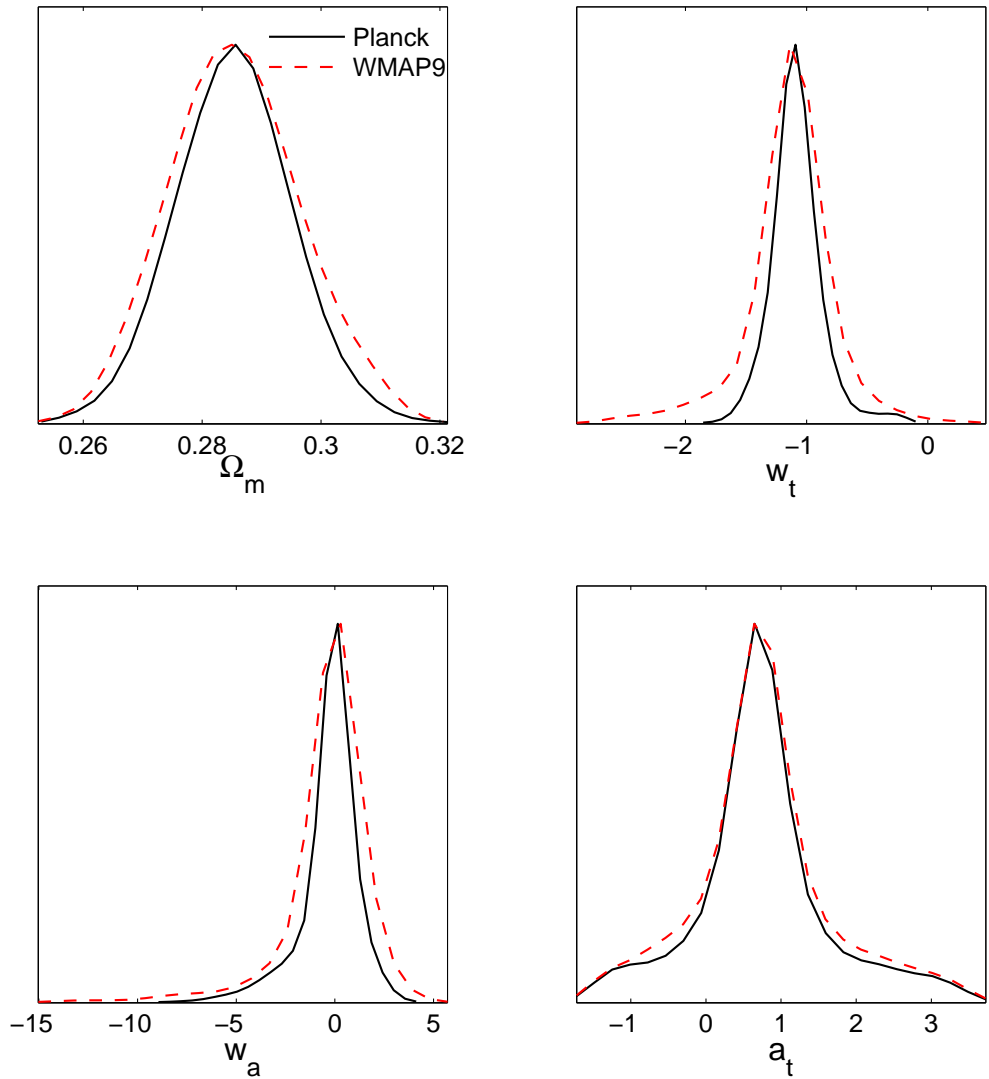}
\end{center}
\caption[]{\small \label{fig1} Marginalized probability distributions of $\Omega_m$, $w_t$, $w_a$ and $a_t$.
In all panels the black lines denote the constraints from the SNLS3+WMAP9+BAO+HST,
and the red lines denote the constraints from the SNLS3+Planck+BAO+HST.
We find a turning in $w$ at $a_t\sim 0.7$ is mildly favored.
The Planck measurement yields to slightly tighter constraints than the WMAP9 measurement.}
\end{figure*}

We combined the above data sets to perform $\chi^2$ analyses.
Since SNLS3, CMB, BAO and HST are effectively independent measurements,
the total $\chi^2$ function is just the sum of all individual $\chi^2$ functions
\begin{equation}\label{chitotal}
\chi^2_{\rm total}=\chi^2_{\rm SNLS3}+\chi^2_{\rm CMB}+\chi^2_{\rm BAO}+\chi^2_{\rm HST}.
\end{equation}
The parabolic model has three dark energy parameters $w_t$, $w_a$ and $a_t$.
Including three other cosmological parameters $\Omega_m$, $\omega_b$ and $h$,
and two nuisance parameters $\alpha$, $\beta$ characterizing the systematic errors of the SNLS3 dataset \citep{SNLS3A},
the full set of free parameters in our analysis is
\begin{equation}\label{Eq:ParSpace}
{\bf P}=\{\Omega_{m},~w_t,~w_a,~a_t,~\omega_b,~h,~\alpha,~\beta\}.
\end{equation}
We modify the public available CosmoMC package~\citep{cosmomc} to explore the parameter space using the Markov Chain Monte Carlo (MCMC) algorithm.
All the parameters listed in Eq. (\ref{Eq:ParSpace}) are fitted simultaneously.
We generate $\mathcal{O}(10^7)$ samples for each set of results presented.

\section{Results}

Table \ref{Table1} summarizes the fitting results of the parabolic model,
including the constraints (mean values and 68\% CL limits) on the main cosmological parameters, and the $\chi^2_{min}$s.
Results from the SNLS3+Planck+BAO+HST (here after Planck combination)
and SNLS3+WMAP9+BAO+HST (hereafter WMAP9 combination) datasets are listed in the 2nd and 3rd columns, respectively.
Figure~\ref{fig1} shows the marginalized probability distributions of $\Omega_m$, $w_t$, $w_a$ and $a_t$.

We find that $\Omega_m$, $w_t$ are well constrained by the current observational data,
while $w_a$ and $a_t$ are marginally constrained to regions (-8,4) and (-1,3).
This supports our statement that a second order expansion of $w(a)$ is sufficiently comprehensive for current observational data.
Results from the Planck combination and WMAP9 combination are consistent with each other,
the only difference being that the Planck combination gives slightly tighter constraints.

Interestingly, the existence of a turning point in $w(a)$ is mildly favored.
The bottom right panel of Fig.~\ref{fig1} shows that the likelihood distribution of $a_t$ has a peak located at $a_t\sim0.7$,
lying within the past epoch $a$=[0,1).
But the sign of turning is only detected at 1$\sigma$ CL --
as listed in Table \ref{Table1},
the 68\% CL constraints from Planck and WMAP9 combinations, $a_t=0.75\pm0.65$ and $0.76^{+0.74}_{-0.75}$, 
have large uncertainties of 87\% and 100\%.

\begin{figure*}[htbp]
\centering
\begin{center}
$\begin{array}{c@{\hspace{0.2in}}c} \multicolumn{1}{l}{\mbox{}} &
\multicolumn{1}{l}{\mbox{}} \\
\includegraphics[scale=0.37]{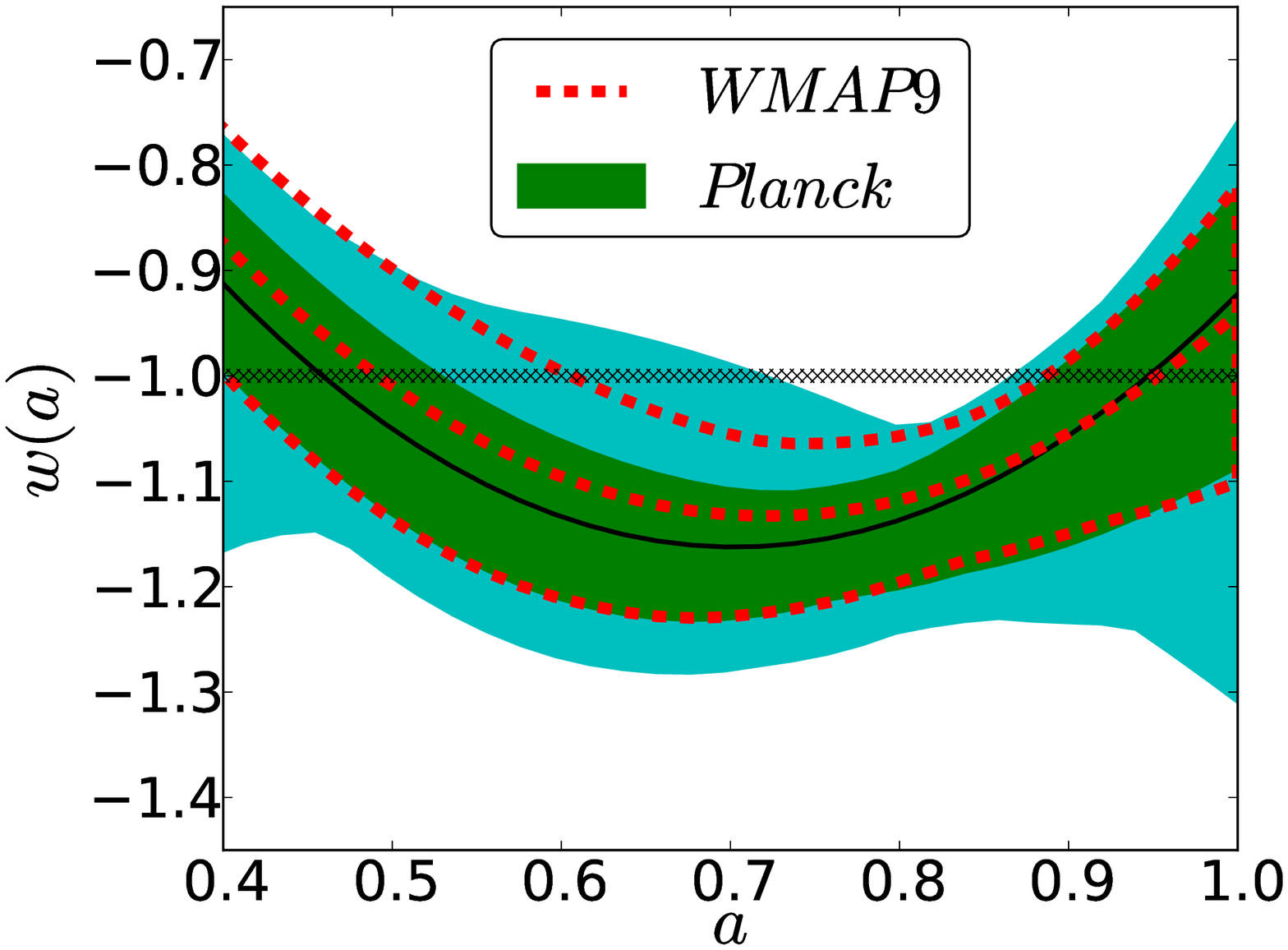} &\includegraphics[scale=0.37]{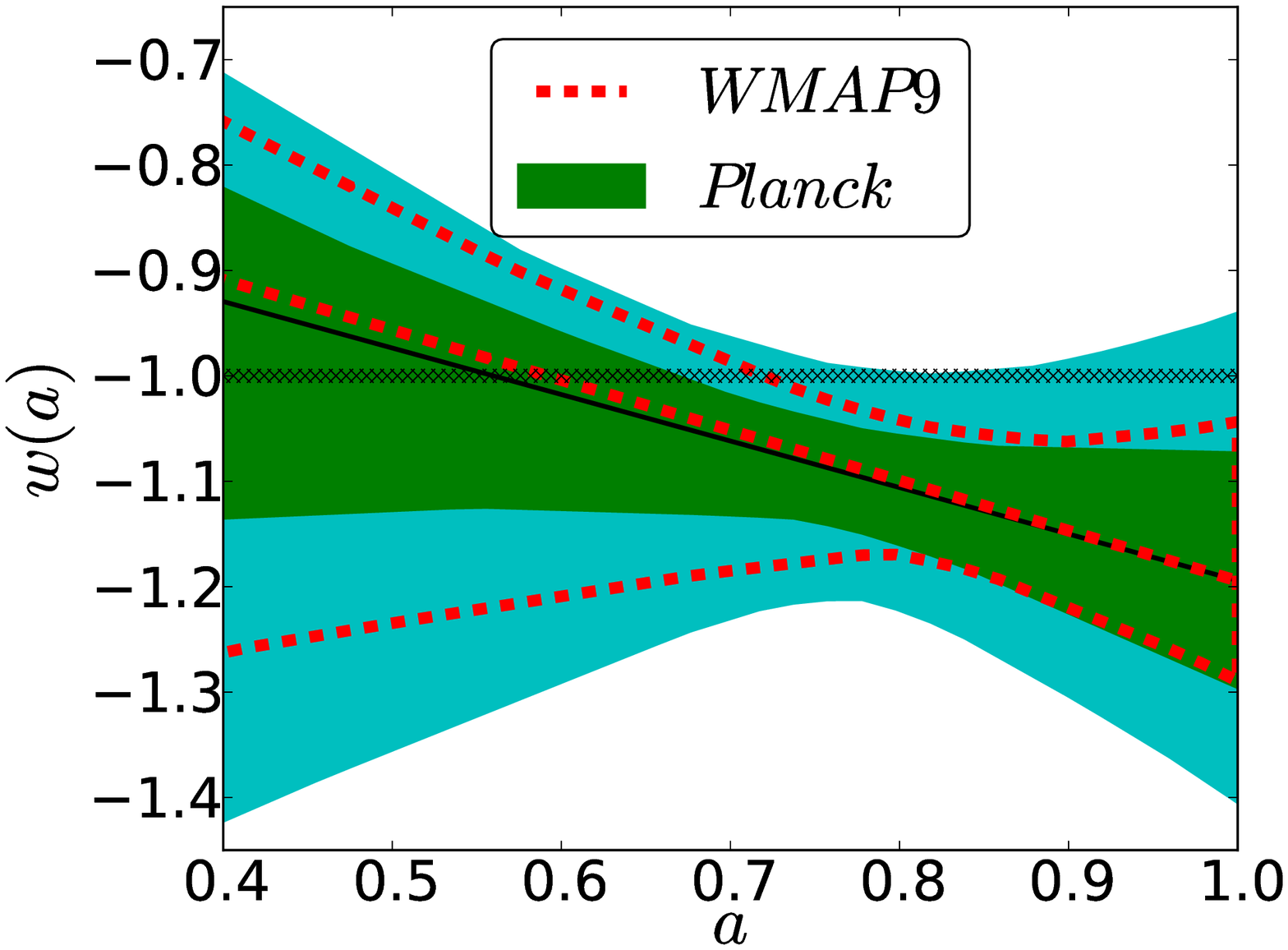} \\
\mbox{(a)} & \mbox{(b)}
\end{array}$
\end{center}
\caption[]{\small \label{fig3}
(a) Reconstructed $w(a)$ as a function of $a$, in the parabolic model.
(b) Reconstructed $w(a)$ in the CPL model.
In both panels, the legend ``WMAP9'' denotes results of the SNLS3+WMAP9+BAO+HST dataset,
with the best-fit and 1$\sigma$ CL regions shown by the red dotted lines,
and the legend ``Planck'' denotes results of the SNLS3+Planck+BAO+HST dataset,
with the best-fit shown by the black solid line and the 1$\sigma$, 2$\sigma$ CL regions shown by filled regions.
$w=-1$ is marked by the horizontal line.
Reconstructed $w(a)$ in the parabolic model has a turning at $a\sim0.7$,
which is not found in the $w(a)$ of the CPL model.}
\end{figure*}

\begin{table*} \caption{$\chi^2_{min}$s, $\chi^2_{min}/{\rm dof}$s and AICs of the $\Lambda$CDM, CPL and parabolic models,
obtained by using the SNLS3+Planck+BAO+HST dataset.}
\begin{center}
\label{Table2}
\begin{tabular}{m{3cm}m{3cm}m{3cm}m{3cm}}
  \hline\hline
  Model                  ~&~$\Lambda$CDM       ~&~CPL               ~&~Parabolic                  \\
  \hline
$\chi^2_{min}$           ~&~431.33             ~&~427.79             ~&~424.83                    \\
$\chi^2_{min}/{\rm dof}$       ~&~0.900              ~&~0.897              ~&~0.893                     \\
$\Delta {\rm AIC}$       ~&~0.00               ~&~0.46               ~&~-0.50                     \\
  \hline\hline
\end{tabular}
\end{center}
\end{table*}

In addition to the cosmological consequence of the parabolic model,
we are also interested in its comparison with other dark energy models.
Thus we also performed the $\chi^2$ analysis of the $\Lambda$CDM and CPL models by using the Planck combination dataset.
To assess different models, we calculated the $\chi^2_{min}/{\rm dof}$ and Akaike Information Criterion (AIC), defined as
\begin{equation}
{\rm dof}=N-k,\ \  {\rm AIC} = \chi^2_{\min}+2k,
\end{equation}
where $N$ is the number of data points, and $k$ is the number of free model parameters.
A model with smaller AIC is more favored.
If the difference of AIC between two models is larger than 2, 
then one model is considered to be favored over the other.

Table~\ref{Table2} lists the $\chi^2_{min}$s, $\chi^2_{min}/{\rm dof}$s and AICs of the $\Lambda$CDM, CPL and parabolic models.
Note that the absolute values of the AICs are not of interest,
thus we list the relative values compared with $\Lambda$CDM, that is, $\Delta {\rm AIC}\equiv {\rm AIC}-{\rm AIC_{\Lambda CDM}}$.

Compared with the $\Lambda$CDM and CPL models,
the parabolic model provides a better fit to the data, reducing the associated $\chi^2_{min}$s by 6.4 and 2.9, respectively.
More interestingly, the parabolic model also gives smallest $\chi^2/{\rm dof}$ and AIC.
The CPL model gives slightly smaller $\chi^2/{\rm dof}$ and larger AIC than $\Lambda$CDM.

Considering the small difference between the AICs,
the performances of the $\Lambda$CDM, CPL and parabolic models are equivalent.
This is different from the result of \cite{Felice2012},
where the authors considered two parametrizations in which $w$ has an extreme,
and found them not favored over the $\Lambda$CDM model under AIC.
This is probably because of the different models and data adopted in the analyses.

In the left panel of Fig.~\ref{fig3}, we plot the reconstructed $w(a)$ of the parabolic model in the range $0.4<a<1$.
The best-fit and 1$\sigma$, 2$\sigma$ regions from the Planck combination are shown by the black line and filled regions.
We find many notable features in the reconstructed $w(a)$:
Within range $0.4<a<1$, the 1$\sigma$ region of $w(a)$ shows a clear shape of turning.
The phantom behavior of dark energy is favored at the extreme point.
At the epoch $0.55\lesssim a\lesssim 0.9$ we find $w<-1$ at 1 $\sigma$ CL,
and at the sweet point $a\approx0.8$, where $w$ is tightest constrained,
$w<-1$ is favored at $2\ \sigma$ CL.
(this is not contradictory to the result $w_t = -1.108^{+0.15}_{-0.16}$ listed in Table (\ref{Table1}),
where $w_t$ is consistent with $w=-1$ at 68\% CL.
The values listed in Table (\ref{Table1}) are marginalized constraints of model parameters;
it is entirely possible that at some $a$ the conditional constraint on $w$ is tighter and different).
In other regions of $a$, the 1$\sigma$ constraint on $w$ is consistent with $w=-1$.
In addition, while the existence of a turning is favored at 1$\sigma$ CL,
it is not clearly detected in the 2$\sigma$ region.

Also, in Fig.~\ref{fig3} we plot the results from the WMAP9 combination.
The best-fit and 1$\sigma$ constraints, shown by the red dashed lines,
are similar to the results from the Planck combination,
except that the WMAP9 combination constraint is slightly weaker.



Fitting results of the CPL model is presented in the right panel of Fig. \ref{fig3} for comparison.
They are consistent with the parabolic model results in that,
a trend that $w$ is evolving from $w\gtrsim-1$ at $a<0.5$ to $w\lesssim-1$ at $a>0.5$ is mildly favored
{similar results have been obtained in \citep{Gongbo Zhao},
where the authors found a dynamical dark energy model
evolving from $w > -1$ at high redshift to $w < -1$ at low redshift is mildly favored),
and the phantom behavior of $w$ at $a\approx0.8$ is favored at 2$\sigma$ CL.
The difference is, the CPL model $w(a)$s, as expected, do not manifest any sign of turning.
This demonstrates that, the parabolic model has advantages in revealing the possible non-linear features in $w$,
and with the progressing of cosmological observations 
considering such high order parametrizations appears to become more and more important.

\section{Conclusion}

We have studied cosmological constraints on the dark energy EOS $w$
by adopting the parabolic parametrization $w(a)=w_t+w_a(a_t-a)^2$,
which is a direct characterization of a possible turning in $w$.
We use the most recent observational data of SNLS3, BAO, CMB and HST to constrain the parameter space of the model.
We found the existence of a turning point in $w$ at $a\sim0.7$ is favored at 1$\sigma$ CL.
$w<-1$ is favored at 1$\sigma$ CL in the epoch $0.55\lesssim a\lesssim 0.9$,
and the significance increases to 2$\sigma$ CL near $a\approx0.8$.
In other regions of $a$, $w$ is consistent with the cosmological constant at 1$\sigma$ CL.
The Planck and WMAP9 combinations give similar fitting results.
The CPL fitting results, although share some common features with the parabolic model results,
do not manifest a turning in $w$.
The $\Lambda$CDM, CPL and parabolic models achieve equivalent performances when we use AIC to assess them.
Our analysis demonstrates that, 
adopting a high order parametrization has the advantages of revealing the possible non-linear features in $w$,
and considering such parametrizations is becoming important currently.


Finally, we mention that a turning in $w$ is only favored at 1$\sigma$ CL.
Based on current observations we are still far from arguing a detection of any dynamical properties or non-linear features in $w$.
Also, our results could be biased because of the form of adopted $w(a)$ and the possible systematic errors in the data.
To reveal the mysterious veil of dark energy, we need to make an exhaustive study based on more powerful data from future experiments.


\begin{acknowledgments}
We would like to thank Cheng Cheng for helpful suggestions.
XDL thanks Seokcheon Lee and Cristiano Sabiu for help offered.
We thank KIAS Center for Advanced Computation for providing computing resources.
ML is supported by the National Natural Science Foundation of China (Grant No. 11275247, and Grant No. 11335012).
XDL is supported by the Korea Dark Energy Search (KDES) grant.
\end{acknowledgments}

\end{document}